\documentclass[a4paper,11pt]{article}
\pdfoutput=1 

\usepackage{jheppub,bm,mathtools,caption,subcaption} 

\usepackage[whole]{bxcjkjatype} 

\usepackage[T1]{fontenc} 
\usepackage[utf8]{inputenc}

\newcommand{\dd}{\mathrm{d}}

\usepackage{tikz}
\usepackage{pgfplots}
\pgfplotsset{compat=1.14}
\usepackage{tikz-3dplot}
\usetikzlibrary{intersections, calc,positioning,decorations.pathreplacing,decorations.pathmorphing,arrows,arrows.meta,patterns,pgfplots.fillbetween,math}

\title{Complexity in a moving mirror model}

\author[]{Yoshiki Sato}

\affiliation[]{Physics Division, National Center for Theoretical Sciences, National Taiwan University, \\
Taipei 10617, Taiwan}

\abstract{
In a two-dimensional conformal field theory with a moving mirror, known as a moving mirror model, the time evolution of the entanglement entropy shows a Page like curve.
This implies that the moving mirror model is useful to understand the island formula.
In this paper, we study the time evolution of the subregion CV complexity in the moving mirror model for a better understanding of the island formula of the complexity.
In contrast to the entanglement entropy, the subregion CV complexity shows a peculiar behaviour.
We discuss this behaviour in more detail.
}

\begin{document} 
\maketitle
\flushbottom

\section{Introduction}

In recent years, significant progress has been made on the information loss problem of black holes.
According to a conventional Hawking's calculation, an entanglement entropy of the Hawking radiation outside the black hole keeps increasing with time \cite{Hawking:1974sw}.
However, if the evaporation process of the black hole is unitary, the entanglement entropy should be zero when the black hole evaporates completely \cite{Page:1993df,Page:1993wv}.
Recently, the formula for the entanglement entropy on a curved geometry, known as the island formula, is proposed \cite{Penington:2019npb,Almheiri:2019psf}
thanks to the recent progress on the holographic entanglement entropy in the AdS/CFT correspondence \cite{Ryu:2006bv,Ryu:2006ef,Lewkowycz:2013nqa,Engelhardt:2014gca}.
The island formula argues that the entanglement entropy of the Hawking radiation in the region $A$ is given by
\begin{align}
\label{island}
    S_\text{rad} (A) = \underset{I}{\text{min}} \left[ \underset{I}{\text{ext}} \left( \frac{\text{Area}(\partial I)}{4G_\text{N}} +S_\text{semi-classical} (A \cup I) \right) \right] \,,
\end{align}
where the disconnected region $I$ is incorporated in the formula, $G_\text{N}$ is the Newton's constant, and $S_\text{semi-classical}(A \cup I)$ is a standard entanglement entropy in quantum field theory.
Using the island formula \eqref{island}, we can calculate the entanglement entropy of the Hawking radiation and confirm that it shows a Page curve.
See, e.g., \cite{Almheiri:2020cfm,Raju:2020smc} for a review on this topic.\footnote{Since there are tons of papers of this topic, our citations are not complete.}
The validity of the island formula is confirmed by several ways: the gravitational path integral in Jackiw-Teitelboim gravity \cite{Penington:2019kki,Almheiri:2019qdq} and so-called double holography \cite{Almheiri:2019hni} in the AdS/BCFT model \cite{Takayanagi:2011zk,Fujita:2011fp,Takayanagi:2020njm}.

The island formula \eqref{island} is proposed as a formula of the entanglement entropy in a  gravitational regime.
We can also consider the island formula of other quantum information quantities.
For example, the island formula of the reflected entropy is studied in \cite{Chandrasekaran:2020qtn,Li:2020ceg}, and that of the capacity of entanglement is studied in \cite{Kawabata:2021hac,Kawabata:2021vyo}.
More interesting quantity is the complexity which measures how many gates are required to prepare a target state from an initial state (See, e.g.,    \cite{Aaronson:2016vto}).
In the context of the AdS/CFT correspondence, there are two different proposals of the holographic dual of the complexity: the ``complexity $=$ volume'' (CV) conjecture  \cite{Susskind:2014,Susskind:2014rva} and the ``complexity $=$ action'' (CA) conjecture \cite{Brown:2015bva,Brown:2015lvg}.
The CV conjecture states that the complexity of a target state on a Cauchy surface $A$ is given by a maximal volume of codimension-one surface anchored to the AdS boundary,
\begin{align}
\label{CVcon}
 C_{\text{V}} (A)= \frac{V(\partial A)}{G_{\text{N}}L_0} \,,
\end{align}
where $L_0$ is the undermined length scale to make the complexity dimensionless.
On the other hand, the CA conjecture states that the complexity is given by the gravitational action on the Wheeler-DeWitt (WDW) patch,
\begin{align}
 C_{\text{A}}(A) = \frac{I_{\text{WDW}}(\partial A)}{\pi \hbar} \,,  
 \label{CAcon}
\end{align}
where $\hbar$ is the Planck constant. 
There is no length scale introduced by hand in the CA conjecture in contrast to the CV conjecture.
The holographic complexity is superior to the holographic entanglement entropy in that the complexity can capture the late time behaviour of a black hole \cite{Susskind:2014moa}.

The holographic complexity is extended to a mixed state, and this is called subregion complexity.
As well as the holographic complexity, there are two proposals of the subregion complexity: the subregion volume complexity \cite{Alishahiha:2015rta} and the subregion action complexity \cite{Alishahiha:2018lfv}.
The subregion CV complexity is defined as the minimal area surrounded by the corresponding Ryu-Takayanagi (RT) surface in a similar way to \eqref{CVcon}.
In \cite{Caceres:2019pgf}, an improvement of the subregion CV complexity is presented.
The subregion CA complexity is defined in a similar way to \eqref{CAcon}.

Recently, the complexity is also discussed in the context of the island formula in several papers \cite{Bhattacharya:2020uun,Hernandez:2020nem,Bhattacharya:2021jrn}. 
In particular, the time evolution of the CV complexity in a double  holographic model \cite{Geng:2020qvw,Geng:2020fxl} is studied in \cite{Bhattacharya:2021jrn}.
The authors consider two types of (subregion) CV complexity: the first one is the subregion CV complexity of the region on an end-of-the-world brane and the second one is the CV complexity of a bath region on the AdS boundary.
It is shown that the renormalized complexity of the subregion CV complexity on the brane increases until the Page time and suddenly becomes zero at the Page time.
The renormalized complexity of the bath region increases forever while it jumps at the page curve. See Fig. 6 and Fig. 9 in \cite{Bhattacharya:2021jrn}.

In this paper, we study the subregion CV complexity in a moving mirror model.
The moving mirror model is a simple and tractable model of the Hawking radiation from black holes, as it consists of quantum field theory with boundary \cite{Davies:1976hi,Birrell:1982ix}.
The recent papers \cite{Akal:2020twv,Akal:2021foz} study entanglement entropies in various models of the moving mirror model and show that the time evolution of the entanglement entropy follows a Page curve.\footnote{See earlier works \cite{Bianchi:2014qua,Hotta:2015huj,Good:2016atu,Chen:2017lum,Good:2019tnf} and recent related works \cite{Kawabata:2021hac,Reyes:2021npy,Ageev:2021ipd} for further study of the moving mirror model.}
The advantage of the moving mirror model is that we can construct various models which mimic an eternal black hole and an evaporating black hole in a simple way.
In addition, we can apply the holography to the moving mirror model for further simplification.
Thus, we can study the time evolution of the holographic complexity in an evaporating black hole like setup.
This is interesting because it is difficult to study the time evolution of a complexity even in a holographic way, i.e. in a double holographic setup.

The organisation of this paper is as follows.
In section \ref{sec2}, we review the moving mirror model and its holographic dual.
After that, we compute the entanglement entropy and confirm that its time evolution follows a Page curve.
In section \ref{sec3}, we compute the subregion CV complexity in the moving mirror model and study its time evolution.
The final section is devoted to conclusion and discussion.
A technical detail is summarized in appendix \ref{app1}.

\section{Moving mirror model and entanglement entropy}
\label{sec2}

We briefly review the moving mirror model in section \ref{sec2.1} and its holographic dual in section \ref{sec2.2}.
After that, we also review a computation of the entanglement entropy in the holographic setup of the moving mirror model.
This section is based on the recent papers \cite{Akal:2020twv,Akal:2021foz}.

\subsection{Moving mirror model}
\label{sec2.1}

We first consider a CFT which lives on $x \geq Z(t)$ region in a Lorentzian flat metric 
\begin{align}
    \dd s^2 & = -\, \dd t^2 + \dd x^2 \\
    & = -\,  \dd u\, \dd v \,,
\end{align}
where we introduce null coordinates
\begin{align}
    u= t-x \,, \qquad v=t+x \,.
\end{align}
The moving mirror is characterised by a trajectory $x=Z(t)$.
Change of variables from $(u,v)$ to $(\tilde{u},\tilde{v})$,
\begin{align}
\label{coord_trans}
    \tilde{u} = p(u) \,, \qquad \tilde{v} = v \,,
\end{align}
maps the metric to
\begin{align}
    \dd s^2 = - \frac{\dd \tilde{u}\, \dd \tilde{v}}{p'(u)} \,.
\end{align}
After a Weyl transformation, the metric becomes flat,
\begin{align}
    \dd s^2 = -\, \dd \tilde{u}\, \dd \tilde{v} = -\, \dd \tilde{t}^2 + \dd \tilde{x}^2 \,, 
\end{align}
where $\tilde{u}=\tilde{t}-\tilde{x},\tilde{v}=\tilde{t}+\tilde{x}$.

We introduce the UV cutoff $\epsilon$ in the $(u,v)$ coordinate to obtain the entanglement entropy.
From \eqref{coord_trans}, this UV cutoff $\epsilon$  is mapped to $\tilde{\epsilon} = \epsilon \sqrt{p'(u)}$ in the $(\tilde{u},\tilde{v})$ coordinate.
Here, the UV cutoff $\tilde{\epsilon}$ is not constant and depends on the $(\tilde{u},\tilde{v})$ coordinate.

From now on, we choose the function $p(u)$ as
\begin{align}\label{2.7}
    t+Z(t) = p \left( t- Z(t) \right)
\end{align}
such that the original CFT is mapped to a CFT with a static mirror located at
\begin{align}
\tilde{u} - \tilde{v} = 0 \,.
\end{align}
Furthermore, we choose that the CFT in the $(\tilde{u},\tilde{v})$ coordinate is in a vacuum state.
That is, the moving mirror model in the $(u,v)$ coordinate is obtained by the conformal transformation \eqref{coord_trans} of BCFT in the $(\tilde{u},\tilde{v})$ coordinate.
The stress tensor in the $(u,v)$ coordinate is given by
\begin{align}
\begin{aligned}
    T_{uu} &= - \frac{c}{24\pi} \{ p(u),u\}= \frac{c}{24\pi} \left( \frac{3}{2} \left( \frac{p''(u)}{p'(u)}\right)^2 - \frac{p'''(u)}{p'(u)} \right)\,, \\
    T_{uv} &= 0 \,, \qquad  T_{vv} = 0 \,,
\end{aligned}
\end{align}
where $\{ p(u),u\}$ is a Schwarzian derivative.

In this paper, we consider two types of the moving mirror model:  escaping mirror and kink mirror.

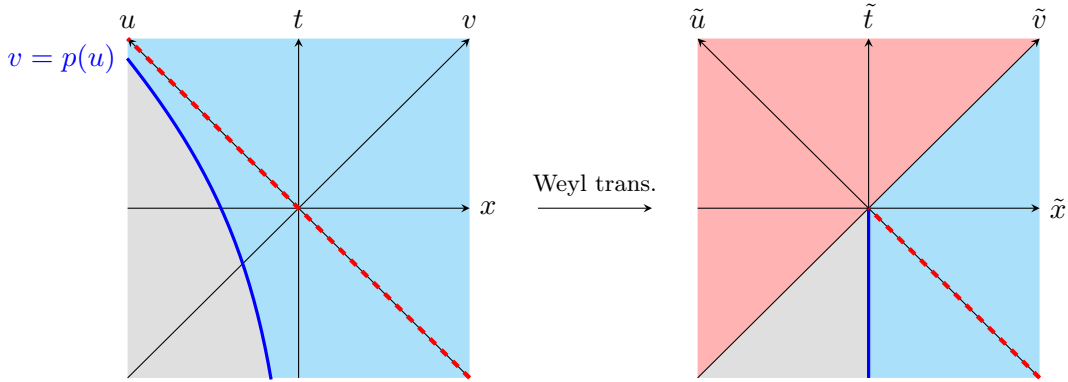
\begin{figure}[ht]
\centering
\begin{tikzpicture}[scale=1.50]

\path[name path=x,domain=-1.5:1.5] plot (\x,-1.5);
\path[name path=p,rotate=-45,domain=0.9:-2] plot(\x,{- ln(1+exp(\x )) });
\path[name path=t,domain=-1.5:1.5] plot (-1.5,\x);
\path node[name intersections={of=x and p}] {};

\tikzmath{
  coordinate \c;
  coordinate \cbase;
  real \px;
      \c1 = (intersection-1);
      \cbase1 = (1,1);
      \px1 = \cx1 / \cbasex1;
      \py1 = \cy1 / \cbasex1;
}

\path node[name intersections={of=t and p}] {};
\tikzmath{
  coordinate \c;
  coordinate \cbase;
  real \px;
      \c2 = (intersection-1);
      \cbase1 = (1,1);
      \px2 = \cx2 / \cbasex1;
      \py2 = \cy2 / \cbasex1;
}

 \fill[lightgray!50!white] (-1.5,-1.5)--(\cx1/1.5,\cy1/1.5)--(-1.5,\cy2/1.5)--cycle;
    \fill[cyan!30!white] (-1.5,\cy2/1.5)--(-1.5,1.5)--(1.5,1.5)--(1.5,-1.5)--(\cx1/1.5,-1.5)--cycle;
    \fill[lightgray!50!white,very thick,domain=0.9:-2,rotate=-45] plot(\x,{- ln(1+exp(\x )) });

    \draw[->,>=stealth] (0,-1.5)--(0,1.5) node [above] {$t$};
    \draw[->,>=stealth] (-1.5,0)--(1.5,0) node [right] {$x$};
    \draw[->,>=stealth] (1.5,-1.5)--(-1.5,1.5) node [above] {$u$};
    \draw[->,>=stealth] (-1.5,-1.5)--(1.5,1.5) node [above] {$v$};
    \draw[blue,very thick,domain=0.9:-2,rotate=-45] plot(\x,{- ln(1+exp(\x )) })node[left]{$v=p(u)$};
    \draw[red,dashed,ultra thick] (1.5,-1.5)--(-1.5,1.5) ;
    
    \draw[->,>=stealth] (2.1,0)--(3.1,0);
    \node at (2.6,0.2)  {\footnotesize{Weyl trans.}}; 
    
    \fill[cyan!30!white] (5,-1.5)--(5,0)--(6.5,1.5)--(6.5,-1.5)--cycle;
    \fill[lightgray!50!white] (3.5,-1.5)--(5,-1.5)--(5,0)--cycle;
    \fill[red!30!white] (3.5,1.5)--(3.5,-1.5)--(6.5,1.5)--cycle;
    
    \draw[->,>=stealth] (5,-1.5)--(5,1.5) node [above] {$\tilde t$};
    \draw[blue,very thick] (5,-1.5)--(5,0);
    \draw[->,>=stealth] (3.5,0)--(6.5,0) node [right] {$\tilde x$};
    \draw[->,>=stealth] (6.5,-1.5)--(3.5,1.5) node [above] {$\tilde u$};
    \draw[->,>=stealth] (3.5,-1.5)--(6.5,1.5) node [above] {$\tilde v$};
    
    \draw[red,dashed,ultra thick] (6.5,-1.5)--(5,0);
\end{tikzpicture}
\caption{A schematic picture of the moving mirror model with the escaping mirror. 
The left figure is written in the $(t,x)$ coordinate, and the right figure is written in the $(\tilde{t},\tilde{x})$ coordinate.
The moving mirror is located at the blue line.
The null line at the boundary of the cyan and pink regions behaves as a black hole horizon.
}
\label{fig:moving_mirror}
\end{figure}

\paragraph{Escaping mirror.}
The escaping mirror is characterised by
\begin{align}
\label{escaping_mirror}
    p(u)= -\beta \log \left(1+\mathrm{e}^{-\frac{u}{\beta}} \right) \,,
\end{align}
with a positive parameter $\beta>0$, which can be regarded as an effective temperature of the radiation.
The non-vanishing component of the stress tensor is
\begin{align}
    T_{uu} = \frac{c}{48 \pi \beta^2} \frac{1+2 \, \mathrm{e}^{-\frac{u}{\beta}}}{\left(1+\mathrm{e}^{-\frac{u}{\beta}} \right)^2} \,.
\end{align}
It vanishes at $u \to - \infty$, while it gradually increases and approaches 
\begin{align}\label{stress_late}
    T_{uu} \simeq \frac{c}{48 \pi \beta^2} 
\end{align}
at $u \to \infty$.
The stress tensor \eqref{stress_late} is similar to a thermal density matrix with temperature $1/2\pi \beta$.
This observation implies that the escaping mirror model mimics eternal black holes.
We draw the escaping mirror model in Fig. \ref{fig:moving_mirror}.\footnote{In the original paper \cite{Akal:2020twv}, its extended version \cite{Akal:2021foz} (except the published version) and the capacity of entanglement paper \cite{Kawabata:2021hac},  similar figures are written. However, the figures in \cite{Akal:2020twv,Kawabata:2021hac,Akal:2021foz} contain a minor correction which is that the trajectory of the moving mirror extends in $\tilde{t} >0 $ region.}
In the $(u,v)$ coordinate, the moving mirror separates the space into a cyan region and a gray region. These regions are mapped to the corresponding colour's regions in the $(\tilde{u},\tilde{v})$ coordinate.
A new pink region, however, appears in the $(\tilde{u},\tilde{v})$ coordinate, and the intersection between the pink and the cyan regions serves as a horizon of a black hole.
An important notice is that the region with $u \to \infty$ and  $v \leq 0$ maps to the origin in the $(\tilde{u},\tilde{v})$ coordinate.

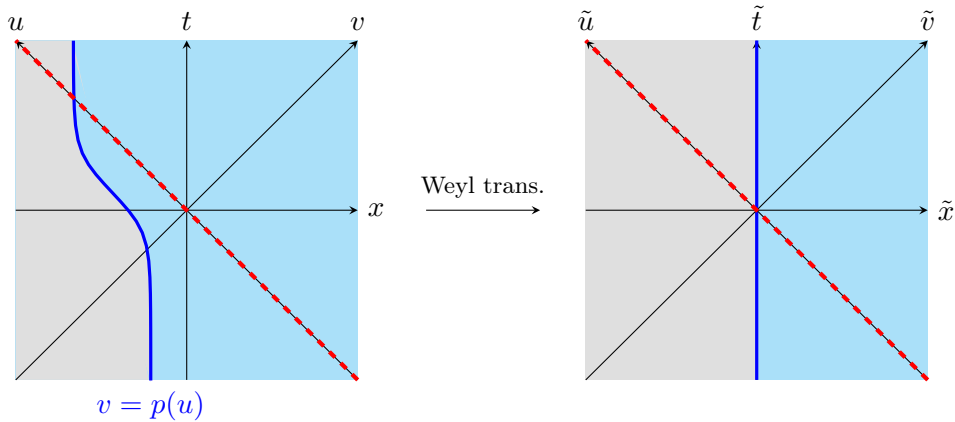
\begin{figure}[ht]
\centering
\begin{tikzpicture}[scale=1.50]

\fill[cyan!30!white] (-1.5,1.5)--(1.5,1.5)--(1.5,-1.5)--(-1.5,-1.5)--cycle;
 \fill[lightgray!50!white] (-1.5,-1.5)--(-1.5,1.5)--(-1.0,1.5)--(-0.33,-1.5)--cycle;
    
    \fill[lightgray!50!white,very thick,domain=-0.3:15.9,rotate=-45,xshift =-30,yshift=-10, scale=0.12] plot(\x  ,{ln(1+exp(-\x  )) - ln(1+exp(\x -8 ))) });
    \fill[cyan!30!white,very thick,domain=-5.3:3.9,rotate=-45,xshift =-30,yshift=-10, scale=0.12] plot(\x  ,{ln(1+exp(-\x  )) - ln(1+exp(\x -8 ))) });

    \draw[->,>=stealth] (0,-1.5)--(0,1.5) node [above] {$t$};
    \draw[->,>=stealth] (-1.5,0)--(1.5,0) node [right] {$x$};
    \draw[->,>=stealth] (1.5,-1.5)--(-1.5,1.5) node [above] {$u$};
    \draw[->,>=stealth] (-1.5,-1.5)--(1.5,1.5) node [above] {$v$};
    \draw[blue,very thick,domain=-5.9:15.8,rotate=-45,xshift =-30,yshift=-10, scale=0.12] plot(\x  ,{ln(1+exp(-\x  )) - ln(1+exp(\x   -8 )))  })node[below]{$v=p(u)$};
    \draw[red,dashed,ultra thick] (1.5,-1.5)--(-1.5,1.5) ;
    
    \draw[->,>=stealth] (2.1,0)--(3.1,0);
    \node at (2.6,0.2)  {\footnotesize{Weyl trans.}}; 
    
    \fill[cyan!30!white] (5,-1.5)--(5,1.5)--(6.5,1.5)--(6.5,-1.5)--cycle;
    \fill[lightgray!50!white] (3.5,-1.5)--(5,-1.5)--(5,1.5)--(3.5,1.5)--cycle;
    
    \draw[->,>=stealth] (5,-1.5)--(5,1.5) node [above] {$\tilde t$};
    \draw[blue,very thick] (5,-1.5)--(5,1.5);
    \draw[->,>=stealth] (3.5,0)--(6.5,0) node [right] {$\tilde x$};
    \draw[->,>=stealth] (6.5,-1.5)--(3.5,1.5) node [above] {$\tilde u$};
    \draw[->,>=stealth] (3.5,-1.5)--(6.5,1.5) node [above] {$\tilde v$};
    
    \draw[red,dashed,ultra thick] (6.5,-1.5)--(3.5,1.5);
\end{tikzpicture}
\caption{A schematic picture of the moving mirror model with the kink mirror. 
The left figure is written in the $(t,x)$ coordinate, and the right figure is written in the $(\tilde{t},\tilde{x})$ coordinate.
The moving mirror is located at the blue line.
}
\label{fig:moving_mirror_kink}
\end{figure}

\paragraph{Kink mirror.}
The kink mirror is characterised by 
\begin{align}\label{kink_mirror}
    p(u) = -\beta \log \left( 1+\mathrm{e}^{-\frac{u}{\beta}} \right) + \beta \log \left( 1+\mathrm{e}^{\frac{u-u_0}{\beta}} \right) \,, 
\end{align}
with a positive $\beta>0$ and $u_0>0$. 
From \eqref{2.7}, one can see that the moving mirror sits at $x=0$ at $t \to - \infty$ and at $x=-u_0/2$ at $t \to \infty$, and the shape of the moving mirror seems a kink.
We draw the kink mirror model in Fig. \ref{fig:moving_mirror_kink}.
In the kink mirror model, a new region does not appear in the $(\tilde{u},\tilde{v})$ coordinate in contrast to the escaping mirror model.

The non-vanishing component of the stress tensor is 
\begin{align}
    T_{uu} = \frac{c}{48 \pi \beta^2} \left( \frac{1+2 \, \mathrm{e}^{-\frac{u}{\beta}}}{\left(1+\mathrm{e}^{-\frac{u}{\beta}}\right)^2} + \frac{1+2 \, \mathrm{e}^{\frac{u-u_0}{\beta}}}{\left(1+\mathrm{e}^{\frac{u-u_0}{\beta}} \right)^2} \right) \,,
\end{align}
and this does not vanish only for a certain period of time.
This observation implies that the kink mirror model mimics evaporating black holes.

\subsection{Holographic dual of the moving mirror model}
\label{sec2.2}

In this section, we review the holographic dual of the moving mirror model based on \cite{Akal:2020twv,Akal:2021foz}.
A more detailed analysis is given in the recent paper \cite{Akal:2021foz}.

Let us consider holographic BCFTs and compute the holographic entanglement entropy.
The holographic dual of BCFTs is known as the AdS/BCFT model \cite{Takayanagi:2011zk,Fujita:2011fp}.
In the AdS$_{d+1}$/BCFT$_d$ model, the gravity dual has a standard AdS$_{d+1}$ metric while the end-of-the-world brane with a tention $\mathcal{T}$ is introduced in the bulk of the AdS$_{d+1}$ spacetime. 
The presence of the brane breaks the symmetry from $SO(d,2)$ to $SO(d-1,2)$.
On the end-of-the-world brane, the Neumann boundary condition is imposed for the induced metric $h_{ab}$,
\begin{align}
    K_{ab} - h_{ab} K + \mathcal{T} h_{ab} = 0 \,, 
\end{align}
where $K_{ab}$ is the extrinsic curvature and $K$ is its trace. 

The metric of the gravity dual in the Poincar\'{e} coordinate is 
\begin{align}
    \dd s^2 = L^2 \frac{\dd \eta^2 - \dd U \dd V }{\eta^2} \,,
\end{align}
where $L$ is the AdS radius, $\eta$ is a radial direction and $U,V$ are null coordinates
\begin{align}
    U = T-X \,, \qquad V = T+X \,.
\end{align}
The end-of-the-world brane extends into the bulk AdS space,  
\begin{align}
    \eta = - \alpha X \,, \qquad \alpha = \frac{\sqrt{1-\mathcal{T}^2}}{\mathcal{T}} \,.
\end{align}
As we will see later, the parameter $\alpha$ is also related to the boundary entropy.

The gravity dual of the original BCFT can be obtained by a coordinate transformation, which is a special case of \cite{Banados:1998gg,Roberts:2012aq},  
\begin{align}
    U = p(u) \,, \qquad V = v +\frac{p''(u)}{2p'(u)} z^2 \,, \qquad \eta = z \sqrt{p'(u)} \,.
\end{align}
Note that $U$ and $V$ coincide with $\tilde{u}$ and $\tilde{v}$ at the boundary of AdS, respectively.
Then, the metric of the $(z,u,v)$ coordinate is given by
\begin{align}
    \dd s^2 = \frac{\dd z^2- \dd u\, \dd v}{z^2} + \frac{12 \pi}{c} T_{uu} \dd u^2 \,.
\end{align}
If the UV cutoff is introduced at $z=\epsilon$, this UV cutoff is mapped to $\eta = \epsilon \sqrt{ p'(p^{-1}(U))}$ which depends on $U$.

\subsection{Entanglement entropy}
\label{sec2.3}

We briefly review a computation of the entanglement entropy of the interval between $x_\text{L}$ and $x_\text{R} (> x_\text{L})$ at time $t$ with the UV cutoff $\epsilon$ in the moving mirror model. 
In this paper, we do not consider a case where the interval is attached to the boundary.
To compute the entanglement entropy, we consider a holographic CFT with large central charge $c$.
See \cite{Akal:2020twv,Kawabata:2021hac,Akal:2021foz} for the detail.
This interval is mapped to the interval 
\begin{align}
\label{interval_tilde}
(\tilde{t},\tilde{x}) = \left( \frac{p(t-x)+t+x}{2}, \frac{-p(t-x)+t+x}{2} \right)    
\end{align}
with $x_\text{L} \leq x \leq x_\text{R}$ in the upper half plane.
The interval in the upper half plane \eqref{interval_tilde} is not a straight line in general.
The entanglement entropy, however, depends on only a causal region.
This means that the entanglement entropy is determined by the end points of the interval.

In the AdS/BCFT model, the left end point of the interval is given by 
\begin{align}
    (T_\text{L},X_\text{L}) &= \left( \frac{p(t-x_\text{L})+t+x_\text{L}}{2}, \frac{-p(t-x_\text{L})+t+x_\text{L}}{2} \right) \,, 
\end{align}
and the right one is given by a similar expression.
We introduce the centre coordinates and the differences of the coordinates
\begin{align}
\begin{aligned}
    X_\text{c} &= \frac{X_\text{R}+X_\text{L}}{2} \,, \quad &
    X_\text{d} &= \frac{X_\text{R}-X_\text{L}}{2} \,, \\
    T_\text{c} &= \frac{T_\text{R}+T_\text{L}}{2} \,, \quad &
    T_\text{d} &= \frac{T_\text{R}-T_\text{L}}{2} \,,
\end{aligned}
\end{align}
for later convenience.
The entanglement entropy of this interval is given by the area of the extremal surface anchored to the interval \cite{Ryu:2006bv},
\begin{align}
    S_A = \frac{\text{Area}(\gamma_A)}{4G_\text{N}} \,,
\end{align}
where $G_\text{N}$ is a Newton's constant.
In the AdS/BCFT model, there are two different extremal surfaces: the connected RT surface and the disconnected RT surface.
The connected RT surface is given by
\begin{align}
    \eta^2 + \left( 1-\left(\frac{T_\text{d}}{X_\text{d}}\right)^2 \right) (X - X_\text{c})^2 = X_\text{d}^2 - T_\text{d}^2 \,, 
\end{align}
where the Lorentz factor appears since the RT surface is tilting in the time direction.
The corresponding holographic entanglement entropy is given by
\begin{align}
    S_A^\text{con} & = \frac{L}{4G_\text{N}} \log \left( \frac{(X_\text{R}-X_\text{L})^2 - (T_\text{R} - T_\text{L})^2}{\epsilon^2 \sqrt{p'(p^{-1}(U_\text{R}))p'(p^{-1}(U_\text{L}))} } \right) \\
    & = \frac{L}{4G_\text{N}} \log \left( \frac{(x_\text{R}-x_\text{L})(p(t-x_\text{L})-p(t-x_\text{R}))}{\epsilon^2 \sqrt{p'(t-x_\text{R} )p'(t-x _\text{L})} } \right) \,, 
\end{align}
where the coefficient is related to the central charge via  $c=3L/2G_\text{N}$.
When the RT surface sits at the same time slice, i.e. $T_\text{d}=0$, the RT surface reduces to a well-known one.
The disconnected RT surface consists of two arcs,
\begin{align}
\begin{aligned}
    &\eta^2 + X^2 = X_\text{L}^2 \,, \qquad \text{at} \quad T= T_\text{L}\,, \\
    &\eta^2 + X^2 = X_\text{R}^2  \,, \qquad \text{at} \quad T= T_\text{R}\,,
\end{aligned}
\end{align}
which anchor to the brane and the AdS boundary.
Note that these arcs sit at different times and the corresponding entanglement wedge is tilting.
Then, the holographic entanglement entropy is 
\begin{align}
    S_A^\text{dis} & = \frac{L}{4G_\text{N}} \log \left( \frac{2X_\text{R}}{\epsilon \sqrt{p'(p^{-1}(U_\text{R}))}} \right) + \frac{L}{4G_\text{N}} \log \left( \frac{2X_\text{L}}{\epsilon \sqrt{p'(p^{-1}(U_\text{L}))}} \right) + S_\text{bdy} \\
    & = \frac{L}{4G_\text{N}} \log \left( \frac{t+x_\text{R}-p(t-x_\text{R})}{\epsilon \sqrt{p'(t-x_\text{R})}} \right) + \frac{L}{4G_\text{N}} \log \left( \frac{t+x_\text{L}-p(t-x_\text{L})}{\epsilon \sqrt{p'(t-x_\text{L})}} \right) + S_\text{bdy} \,, 
\end{align}
with the boundary entropy, $S_\text{bdy}=(L/4G_\text{N})\text{arcsinh} \, \alpha$.
In total, the holographic entanglement entropy is given by a minimal one
\begin{align}
    S_A = \text{min} (S_A^\text{con}, 
    S_A^\text{dis}) \,.
\end{align}
In the following, we compute the time evolution of the holographic entanglement entropy in the escaping mirror model and the kink mirror model.

\paragraph{Escaping mirror.}

\begin{figure}[th!]
    \centering
    \includegraphics[width=7.5cm]{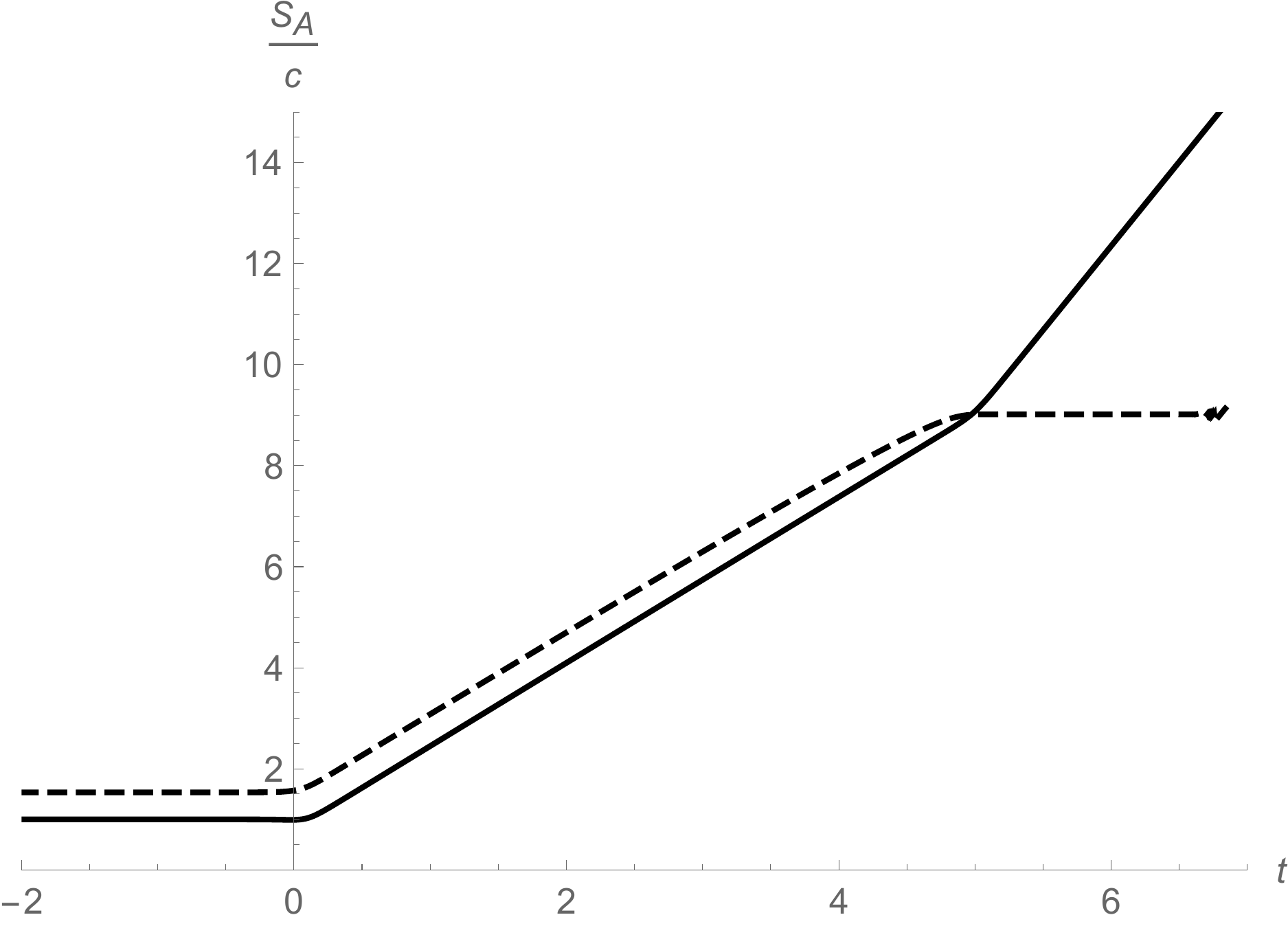}
    \caption{Time evolution of the entanglement entropy of the interval $A = [Z(t)+0.1,Z(t)+10]$ in the escaping mirror model. The parameters are $\beta = 0.1,\epsilon = 0.1$. The solid line and the dashed line correspond to the disconnected entanglement entropy and the connected entanglement entropy, respectively.}
    \label{fig:ee_eternal}
\end{figure}

The time evolution of the entanglement entropy is plotted in Fig. \ref{fig:ee_eternal}.
We choose the region $A=[Z(t)+0.1,Z(t)+10]$ and the parameter as $\beta=0.1,\epsilon = 0.1$ for numerics.
We also drop the boundary entropy of the disconnected phase to see the  logarithmic divergent behaviour.
At an early time, the disconnected phase of the entanglement entropy, depicted in solid line, is favoured. 
At a late time, the entanglement entropy of the connected phase, depicted in the dashed line, becomes smaller than that of the disconnected phase.
This behaviour implies that the escaping mirror model imitates eternal black holes.

\paragraph{Kink mirror.}

\begin{figure}[th!]
    \centering
    \includegraphics[width=7.5cm]{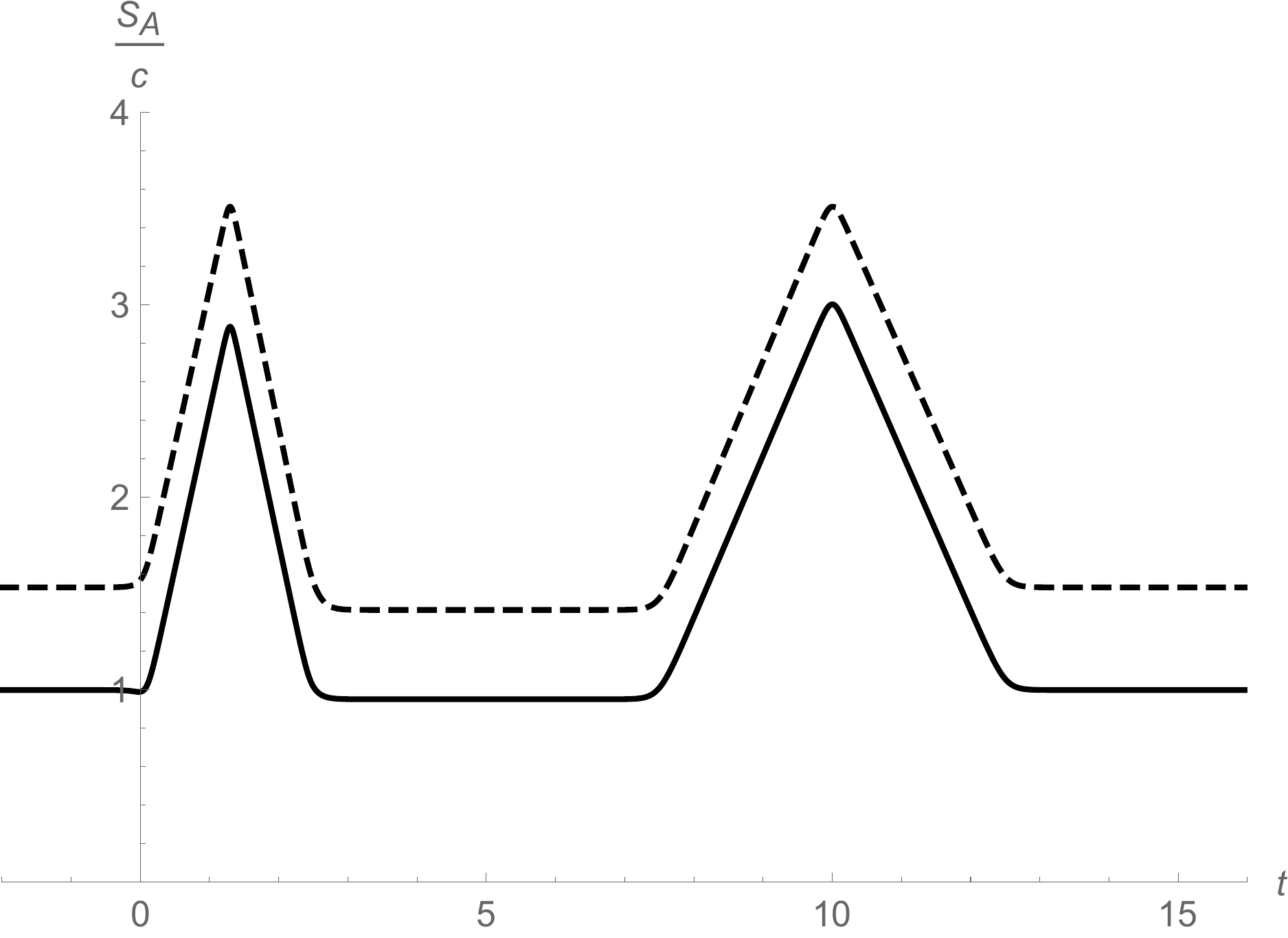}
    \caption{Time evolution of the entanglement entropy of the interval $A = [Z(t)+0.1,Z(t)+10]$ in the kink mirror model. The parameters are $\beta = 0.1,\epsilon = 0.1, u_0 = 5$. The solid line and the dashed line correspond to the disconnected entanglement entropy and the connected entanglement entropy, respectively.}
    \label{fig:ee_eva}
\end{figure}

The time evolution of the entanglement entropy is plotted in Fig. \ref{fig:ee_eva}.
We again choose the region $A=[Z(t)+0.1,Z(t)+10]$ and the parameter as $\beta=0.1,\epsilon = 0.1, u_0 =5$.
We also drop the boundary entropy of the disconnected phase to see the logarithmic divergent behaviour.
The entanglement entropy of the disconnected phase, depicted in the solid line, is always smaller than that of the connected phase, depicted in the dashed line.
Nevertheless, the entanglement entropy shows a Page like curve.
This result is surprising because a Page curve of a double holography setup is realised as a phase transition of the entanglement entropy from a connected phase to a disconnected phase in general.
The above observation implies that the kink mirror model mimics evaporating black holes apart from the point that the entanglement entropy of the kink mirror model has two mountains in contrast to that of evaporating black holes.

\section{Complexity}
\label{sec3}

In this section, we would like to compute the subregion CV complexity in the moving mirror model.
For holographic CFTs, it is possible to compute the subregion complexity using the subregion CV formula or the subregion CA formula.
We study only the CV formula in this paper because it is technically difficult to compute the CA complexity.
The CV complexity in the AdS/BCFT model has been investigated in \cite{Sato:2019kik,Braccia:2019xxi} and the subregion one has been investigated in \cite{Braccia:2019xxi}.\footnote{The complexity in AdS/DCFT models is discussed in \cite{Chapman:2018bqj,Auzzi:2021nrj,Baiguera:2021cba}.}
However, we are interested in the subregion CV complexity which does not sit at the constant time slice, and such a situation is not treated in \cite{Braccia:2019xxi}.
To the best of our knowledge, the (subregion) CV complexity of the tilting region has not investigated.

We are interested in the subregion CV complexity of the interval between $(t,x_\text{L})$ and $(t,x_\text{R})$ with $x_\text{R} > x_\text{L}$ in the original CFT.
Again, we do not consider a case where the interval is attached to the boundary of BCFTs.
The region $A$ at a constant time slice in the $(t,x)$ coordinate is mapped to a curved interval $\tilde{A}$ in the $(\tilde{t},\tilde{x})$ coordinate in general.
However, it is difficult to compute the subregion CV complexity of the curved interval.
Instead of computing its complexity, we compute the subregion CV complexity of the straight interval whose endpoints are the same as the curved interval.
That is, we compute the subregion CV complexity of the straight line given by
\begin{align}
    T - T_{\text{L}} = \frac{T_\text{R}-T_\text{L}}{X_\text{R}-X_\text{L}} (X-X_\text{L}) \,,
\end{align}
where $X$ runs from $X_\text{L}$ to $X_\text{R}$.
We denote this straight line by $\tilde{A}'$ and the corresponding curved line in the $(t,x)$ coordinate by $A'$ to distinguish the curved line $\tilde{A}$ in the $(\tilde{t},\tilde{x})$ coordinate which is mapped from the straight line $A$.

Before computing the complexity, we discuss the causality dependence of the entanglement entropy and the (subregion) complexity in more detail.
In quantum field theory, a reduced density matrix of the region $A$ is obtained by tracing out the region outside of $A$ in a Cauchy surface.
The reduced density matrices of different Cauchy surfaces with the same boundary obtained in this way are connected by a unitary matrix.
Thus, the entanglement entropy does not depend on the choice of the Cauchy surface,
\begin{align}
    S_A = S_{A'} \,,
\end{align}
where $A$ and $A'$ are related by causality.
This is apparent from the holographic viewpoint as the holographic entanglement entropy depends only on the Ryu-Takayanagi surface.
On the other hand, the complexity is defined as the number of required gates to prepare a target state from a reference state.
Thus, it is expected that the complexities are different by a choice of the Cauchy surfaces,
\begin{align}
    C(A) \neq C(A') \,.
\end{align}
In the context of holography, this is apparent from the definition of the (subregion) CV and CA complexities.

Let us move to a discussion of the subregion CV complexity of the region $\tilde{A}'$.
Since $\tilde{A}'$ tilts, it is difficult to find extremal surfaces anchored to $\tilde{A}'$.
Hence, we obtain the extremal surfaces by using a fact that they can be obtained by Lorentz transformation of extremal surfaces in a constant time.
According to the shape of the entanglement wedge, there are two phases of the subregion CV complexity.
The subregion CV complexity of the connected phase is given by
\begin{align}
\begin{aligned}
    C_\text{V}(\tilde{A}') &= \frac{1}{4G_\text{N}L_0} \int_{X_\text{L}}^{X_\text{R}} \! \dd X \int_{\epsilon \sqrt{p'(p^{-1}(U))}}^{\sqrt{1-(T_\text{d}/X_\text{d})^2}\sqrt{X_\text{d}^2-(X-X_\text{c})^2}} \! \dd \eta \,  \frac{L^2}{\eta^2}  \sqrt{1-\left(\frac{T_\text{d}}{X_\text{d}}\right)^2} \\
    & = \frac{L^2}{4G_\text{N}L_0} \left(  \sqrt{1-\left(\frac{T_\text{d}}{X_\text{d}}\right)^2} \int_{X_\text{L}}^{X_\text{R}} \frac{\dd X}{\epsilon \sqrt{p'(p^{-1}(U))}} - \pi \right) \,,
    \label{CV_connected}    
\end{aligned}
\end{align}
where 
\begin{align}
    U = T - X = \frac{T_\text{R}-T_\text{L}}{X_\text{R}-X_\text{L}} (X-X_\text{L}) + T_\text{L} - X \,.
\end{align}
The Lorentz factor appears since the interval is tilting.
The universal part is always negative.
The subregion CV complexity of the disconnected phase is given by
\begin{align}
    C_\text{V}(\tilde{A}') = \frac{L^2}{4G_\text{N}L_0} \left( \sqrt{1-\left(\frac{T_\text{d}}{X_\text{d}}\right)^2} \int_{X_\text{L}}^{X_\text{R}} \frac{\dd X}{\epsilon \sqrt{p'(p^{-1}(U))}} + \alpha \log \left( \frac{X_\text{R}}{X_\text{L}} \right) \right) \,.
    \label{CV_disconnected}
\end{align}
The derivation is devoted in appendix \ref{app1}.
Both CV complexities show the same divergent structures,
\begin{align}
    C_\text{V} (t) & = \frac{C_1 (t)}{\epsilon} + C_0 (t) \,, \\
    C_1 (t) & = \frac{L^2}{4G_\text{N}L_0} \sqrt{1-\left(\frac{T_\text{d}}{X_\text{d}}\right)^2} \int_{X_\text{L}}^{X_\text{R}} \frac{\dd X}{\sqrt{p'(p^{-1}(U))}} \,, \label{c1t} \\
    C_0 (t) & = \frac{L^2}{4G_\text{N}L_0} \times
    \begin{dcases}
    - \pi & \text{connected phase}\,, \\
    \alpha \log \left( \frac{X_\text{R}}{X_\text{L}} \right) & \text{disconnected phase} \,.
    \end{dcases}
    \label{c0t}
\end{align}
Here we write the time dependence explicitly instead of $\tilde{A}'$.
Roughly speaking, $C_1(t)$ expresses a geometric structure of the partial region rather than quantum aspects of the state, and $C_0(t)$ expresses a complexity of the state. We will see this in the following examples.
Since the coefficients of the divergent term are the same in \eqref{CV_connected} and \eqref{CV_disconnected}, the difference of the complexities comes from the finite parts, which does not depend on the choice of the cutoff $\epsilon$.
The finite terms do not depend on the tilt of the entanglement wedge.
As long as the parameter $\alpha$ is not negative, the finite part of the disconnected phase is always larger than that of the connected phase.

In the following, we study the subregion CV complexity of the escaping mirror model and the kink mirror model.

\paragraph{Escaping mirror.}
As shown in \cite{Akal:2020twv,Kawabata:2021hac,Akal:2021foz} and reviewed in section \ref{sec2.3}, the phase of the entanglement wedge changes from the disconnected phase to the connected phase in the time evolution, and this phase transition shows a Page curve like eternal black holes.
Thus, the subregion CV complexity also changes from the disconnected phase to the connected phase, but the divergent parts of the complexities are the same.

\begin{figure}[th]
  \begin{minipage}[b]{0.5\linewidth}
    \centering
    \includegraphics[width=6.5cm]{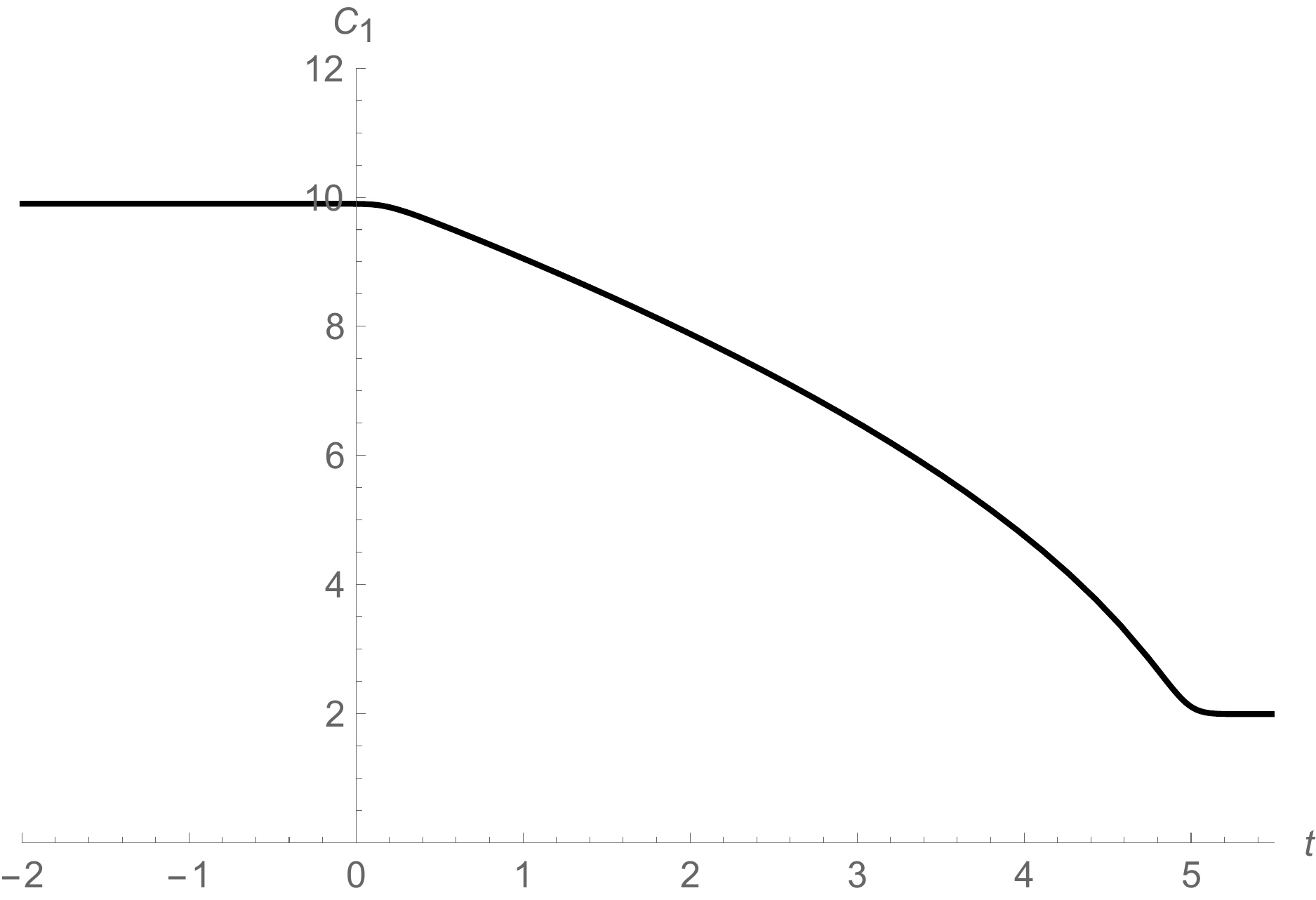}
    \subcaption{$C_1(t)$}
    \label{fig:cv_eternal}
  \end{minipage}
  \begin{minipage}[b]{0.5\linewidth}
    \centering
    \includegraphics[width=6.5cm]{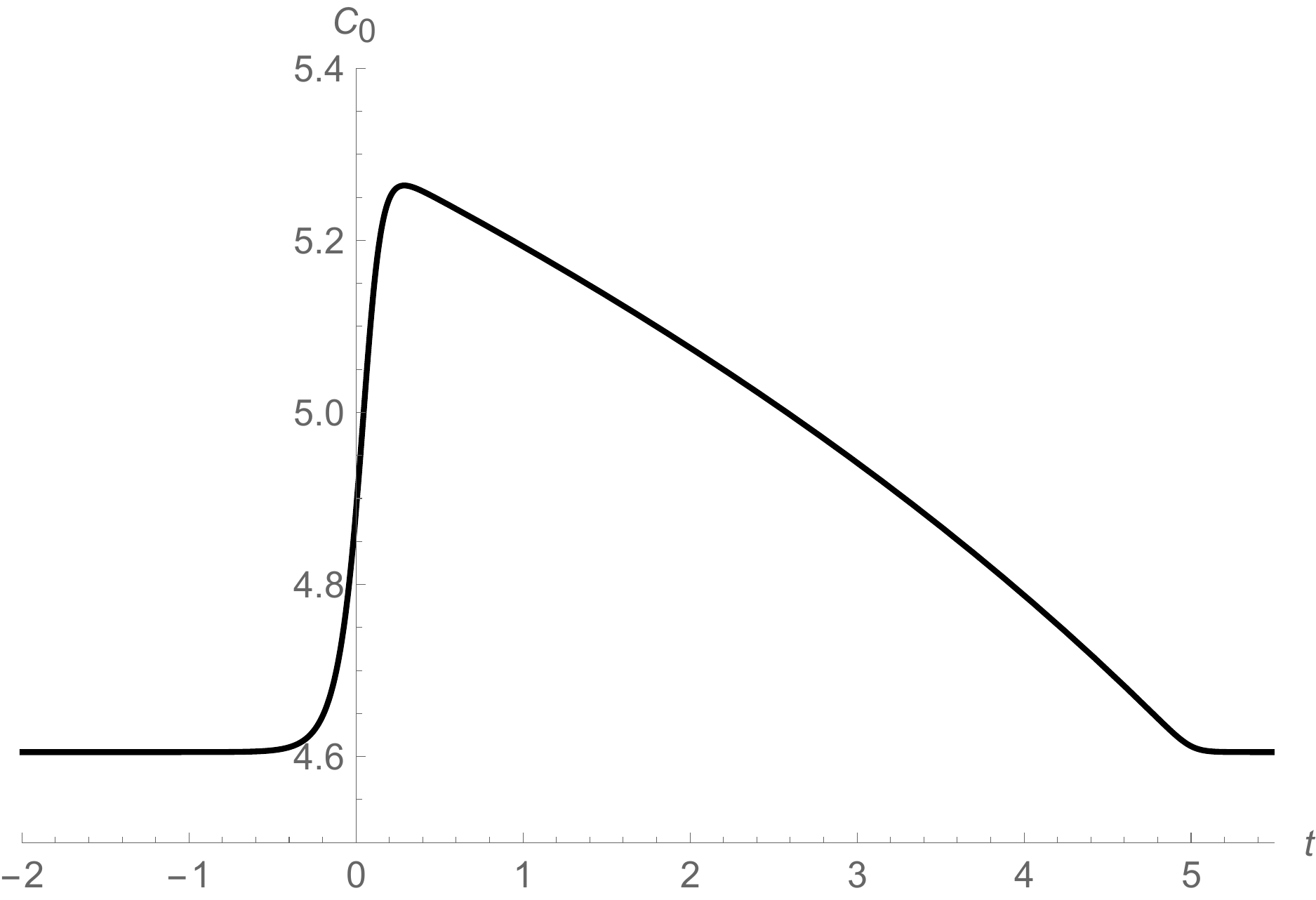}
    \subcaption{$C_0(t)$}
    \label{fig:const_eternal}
  \end{minipage}
  \caption{Time evolution of $C_1(t)$ and $C_0(t)$ of the escaping mirror model. The region is a curved interval between $(t,Z(t)+0.1)$ and $(t,Z(t)+10)$ in the $(t,x)$ coordinate such that the interval is mapped to a straight line in the $(\tilde{t},\tilde{x})$ coordinate.
The parameters are $\beta=0.1$,  $L^2/4G_\text{N}L_0=1$ and $\alpha=1$.}
\end{figure}

\begin{figure}[th]
    \centering
    \includegraphics[width=7.5cm]{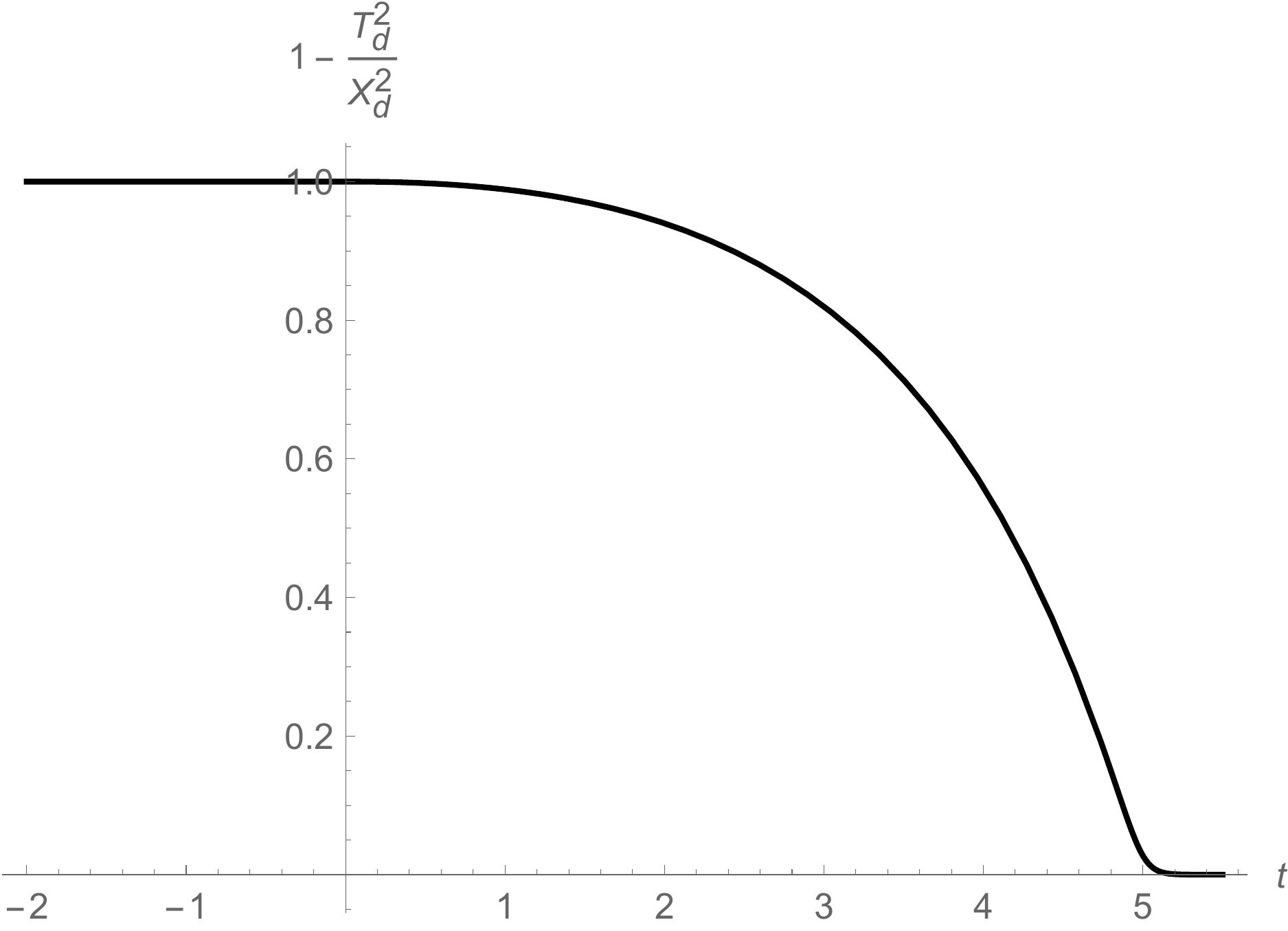}
    \caption{Time evolution of the Lorentz factor of the escaping mirror model.
    The parameters are $\beta=0.1$ and   $L^2/4G_\text{N}L_0=1$.}
    \label{fig:lorentz_eternal}
\end{figure}

We choose the region $A'$ whose endpoints are $(t,Z(t)+0.1)$ and $(t,Z(t)+10)$ in the $(t,x)$ coordinate such that the interval is mapped to a straight line $\tilde{A}'$ in the $(\tilde{t},\tilde{x})$ coordinate.
For numerics, we choose the parameters $\beta=0.1$, $L^2/4G_\text{N}L_0=1$ and $\alpha=1$.
We plot $C_1(t)$ of the subregion CV complexity \eqref{c1t} in Fig. \ref{fig:cv_eternal} and $C_0(t)$ of the disconnected phase of the subregion CV complexity \eqref{c0t} in Fig. \ref{fig:const_eternal}. 
We also plot the time evolution of the Lorentz factor in Fig. \ref{fig:lorentz_eternal}, and we find that the interval becomes null at late time.

Fig. \ref{fig:cv_eternal} shows that $C_1(t)$ decreases until a point ($t \sim 5.0$) where the phase transition of the entanglement entropy occurs.
The decreasing behaviour for $0<t<5$ seems peculiar since the state keeps absorbing Hawking radiation, and it is expected that the cost to prepare the state becomes large.
The decreasing behaviour can be understood as follows.
The interval of $\tilde{A}'$ gradually becomes null since the left endpoint of $\tilde{A}'$ is very close to the origin in the $(\tilde{t},\tilde{x})$ coordinate and the right endpoint moves upward in the time evolution.
Roughly speaking, $C_1(t)$ gives a length of the interval $\tilde{A}'$, and it decreases as in Fig. \ref{fig:cv_eternal}.
Fig. \ref{fig:const_eternal} shows that the time evolution of $C_0(t)$ of the disconnected phase.
As we saw in Fig. \ref{fig:ee_eternal}, the transition from the disconnected phase to the connected phase happens around $t=5$.
Then, $C_0(t)$ jumps from $4.6$ to $-\pi$ suddenly at the transition point. 
This transition represents a Page like curve of the subregion complexity.

\paragraph{Kink mirror.}

\begin{figure}[th]
  \begin{minipage}[b]{0.5\linewidth}
    \centering
    \includegraphics[width=6.5cm]{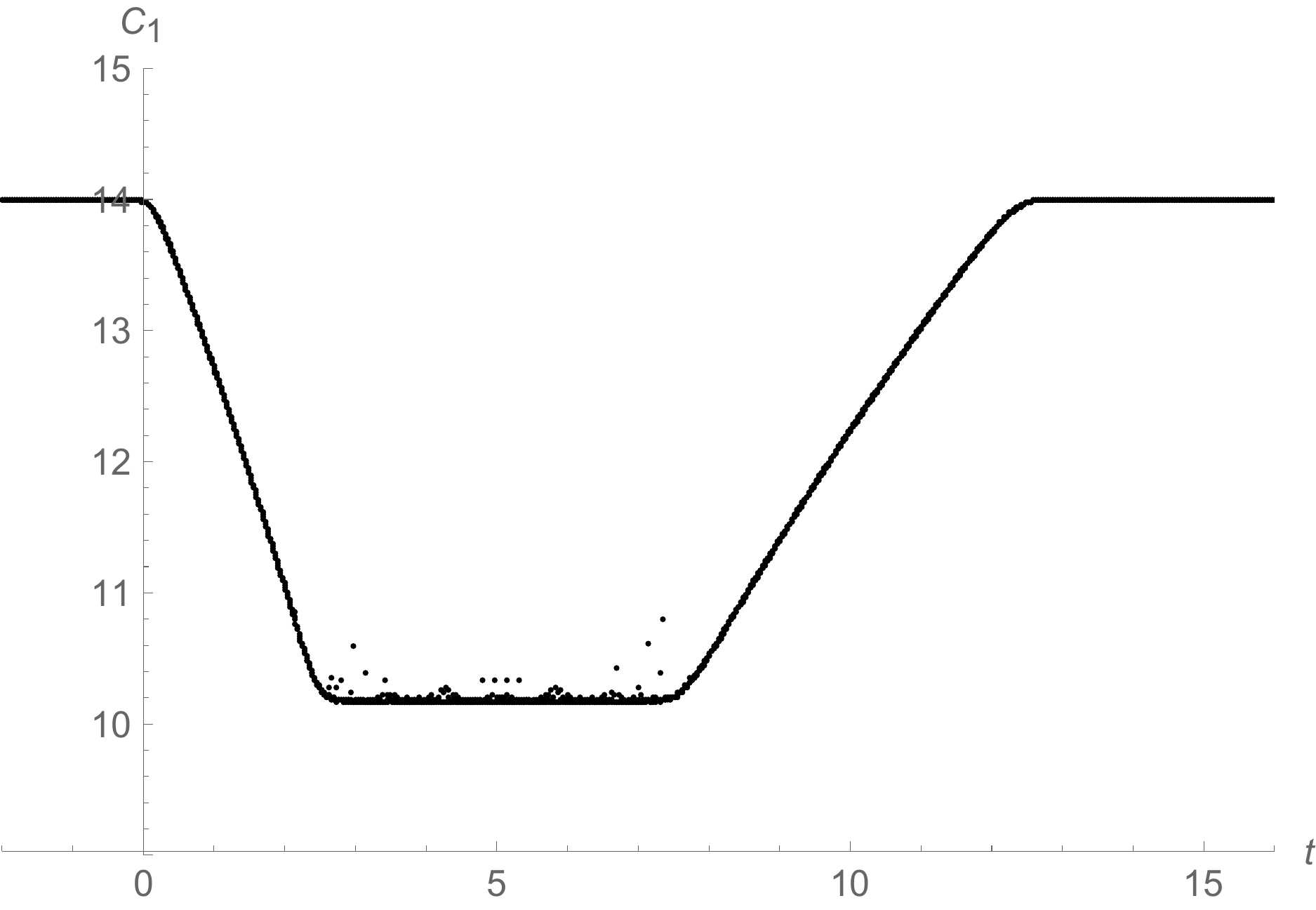}
    \subcaption{$C_1(t)$}
    \label{fig:cv_eva}
  \end{minipage}
  \begin{minipage}[b]{0.5\linewidth}
    \centering
    \includegraphics[width=6.5cm]{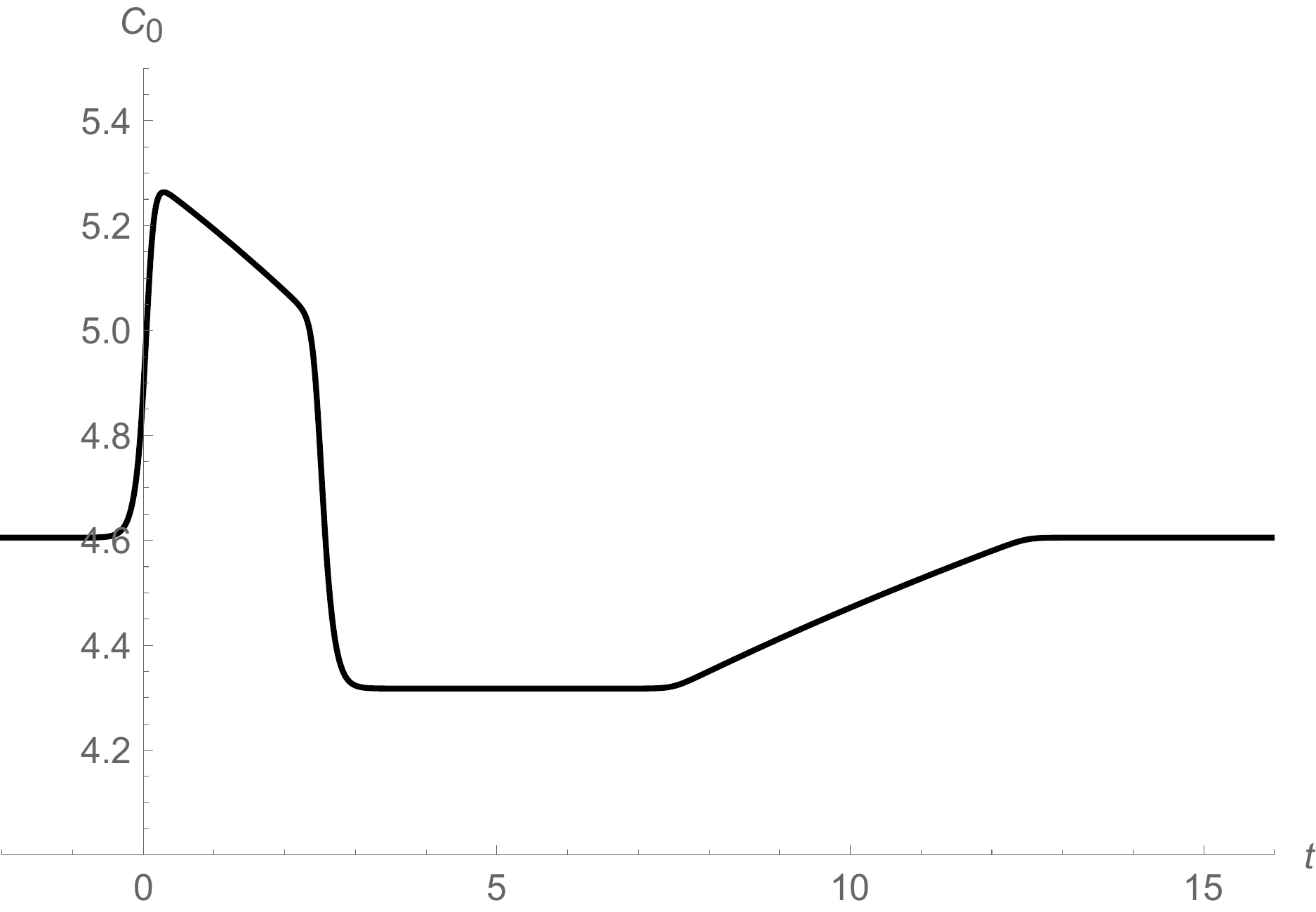}
    \subcaption{$C_0(t)$}
    \label{fig:const_eva}
  \end{minipage}
  \caption{Time evolution of $C_1(t)$ and $C_0(t)$ of the kink model.
    The region is a curved interval between $(t,Z(t)+0.1)$ and $(t,Z(t)+10)$ in the $(t,x)$ coordinate such that the interval is mapped to a straight line in $(\tilde{t},\tilde{x})$ coordinate. The parameters are $\beta=0.1$,  $L^2/4G_\text{N}L_0=1$, $u_0 =5$ and $\alpha =1$. Due to numerical errors, there are many dots around the bottom in Fig. \ref{fig:cv_eva}.}
\end{figure}

\begin{figure}[th]
    \centering
    \includegraphics[width=7.5cm]{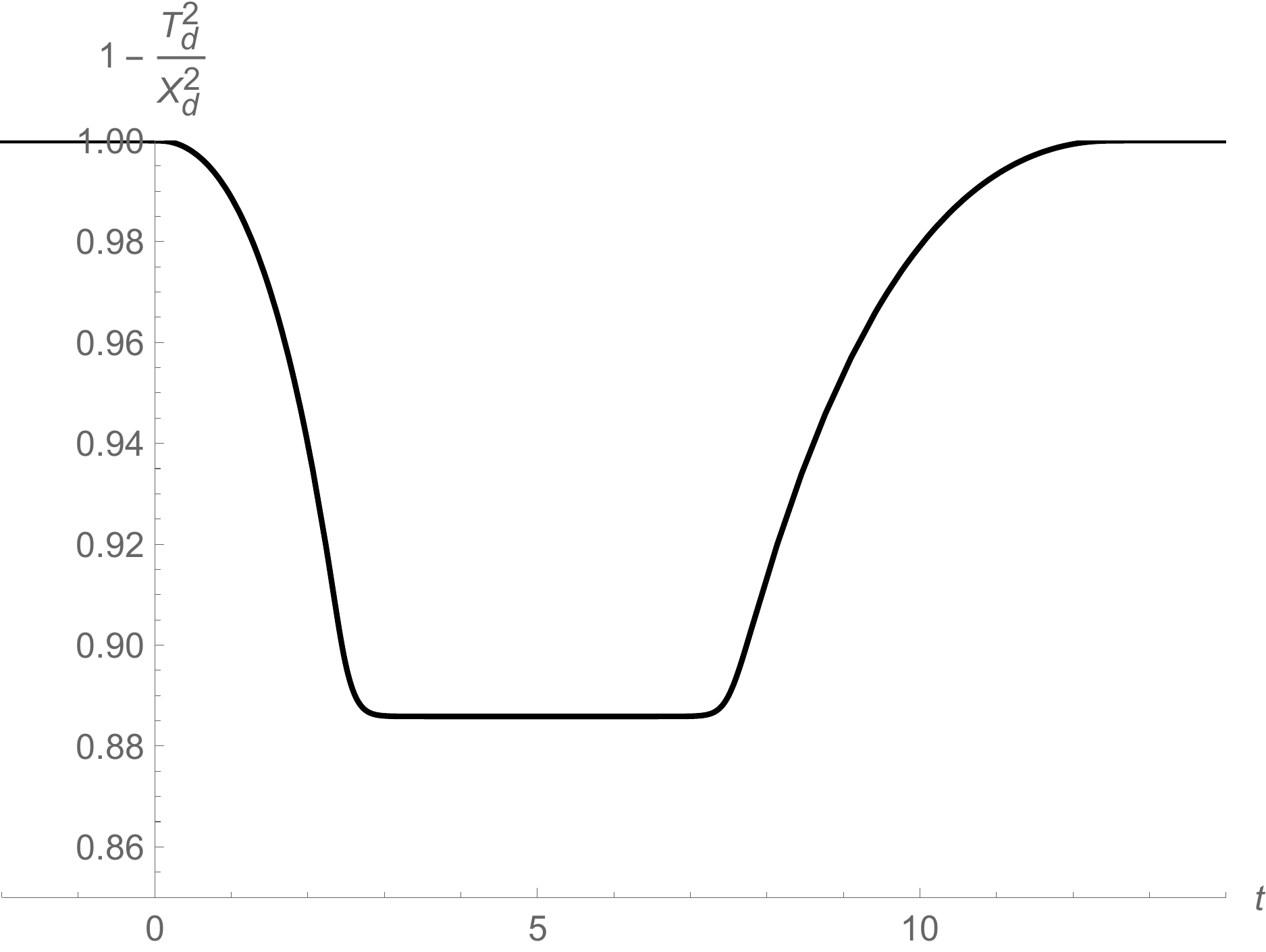}
    \caption{Time evolution of the Lorentz factor of the kink mirror model.
    The parameters are $\beta=0.1$, $L^2/4G_\text{N}L_0=1$ and $u_0=5$.}
    \label{fig:lorentz_eva}
\end{figure}

In the kink mirror model, the disconnected phase is always favoured as seen in section \ref{sec2.3}.
For a numerical computation, we again choose the region $A'$ whose endpoints are $(t,Z(t)+0.1)$ and $(t,Z(t)+10)$ in the $(t,x)$ coordinate such that the interval is mapped to a straight line $\tilde{A}'$ in the $(\tilde{t},\tilde{x})$ coordinate.
We also choose the parameters $\beta=0.1$, $L^2/4G_\text{N}L_0=1$ and $u_0=5$.
We plot $C_1(t)$ of the subregion CV complexity \eqref{c1t} in Fig. \ref{fig:cv_eva}, which contains numerical errors, and $C_0(t)$ of the disconnected phase of the subregion CV complexity \eqref{c0t} in Fig. \ref{fig:const_eva}. 
We also plot the time evolution of the Lorentz factor in Fig. \ref{fig:lorentz_eva}.

$C_1(t)$ decreases during $0<t<2.5$, becomes constant during $2.5 < t < 7.5$ and increases until $t=12.5$.
This behaviour can be understood as the combination of the escaping mirror model and the time-reversed escaping mirror model.
To compare Fig. \ref{fig:ee_eva} with Fig. \ref{fig:cv_eva}, we find that the complexity changes when the entanglement entropy changes.
From the entanglement entropy in Fig. \ref{fig:ee_eva}, the difference of two mountains is not obvious.
However, the complexity makes the difference explicit.

Fig. \ref{fig:const_eva} shows the time evolution of $C_0(t)$ of the disconnected phase which is always favoured as we saw in Fig. \ref{fig:ee_eva}.
In contrast to the previous example, the finite part of the complexity does not show any jump.
$C_0(t)$ irregularly changes according to the movement of the mirror, and returns to the same value finally.
This is a Page like curve of the subregion complexity in the kink mirror model.

\section{Discussion}

We discussed the time evolution of the subregion CV complexity in a moving mirror model for a better understanding of the island formula of the complexity.
In particular, we studied two models, the escaping mirror model which mimics eternal black holes and the kink mirror model which mimics evaporating black holes, and we focused on the coefficients of the divergent part \eqref{c1t} and the finite part of the complexity \eqref{c0t}.
The divergent part are the same both in the connected phase and the disconnected phase.
We numerically obtained their time evolution and plotted in Figs. \ref{fig:cv_eternal}, \ref{fig:const_eternal}, \ref{fig:cv_eva} and \ref{fig:const_eva}. 

Contrary to expectations, the holographic complexity in the moving mirror model sometimes decreases in time evolution (See Fig. \ref{fig:cv_eternal} and Fig. \ref{fig:cv_eva} for the divergent part and Fig. \ref{fig:const_eternal} and Fig. \ref{fig:const_eva} for the finite part).
Naively, it is expected that the complexity increases when the state absorbs the Hawking radiation because it seems that more gates are needed to prepare such a state from an initial state.
The divergent parts, $C_1(t)$, are roughly the length of the region and mainly represent the geometric structures of quantum states.
On the other hand, the finite parts, $C_0(t)$, seem  to represent the quantum aspects of the state.
Hence, the decreasing behaviour of $C_1(t)$ is not peculiar.
$C_0(t)$ of the escaping mirror model shows a transition at the same point where the phase transition of the entanglement entropy occurs.
This behaviour is consistent with the observation in \cite{Bhattacharya:2021jrn}.
$C_0(t)$ of the kink mirror model shows a peculiar result, and a further study on the complexity of evaporating black holes is needed. 

The holographic dual of the moving mirror model does not contain a black hole in the bulk.
The linear growth of the holographic complexity comes from the interior of the black hole.
Hence, our study reveals that the (subregion) complexity can detect the existence of black hole in contrast to the  entanglement entropy.

In this paper, we consider two moving mirror model: the escaping mirror model and the kink mirror model.
We can also consider more general shapes of the moving mirror.
For simplicity, we consider two situations, (i) the moving mirror is not across the $v=0$ curve as the escaping mirror model, and (ii) the moving mirror is across the $v=0$ curve as the kink mirror model.
In the case (i), we will obtain an entanglement entropy and a subregion complexity quantitatively similar to those of the escaping mirror model because we would obtain a picture similar to the right figure in Fig. \ref{fig:moving_mirror}.
In the case (ii), we will obtain an entanglement entropy and a subregion complexity quantitatively similar to those of the kink mirror model because we would obtain a picture similar to the right figure in Fig. \ref{fig:moving_mirror_kink}.

We studied the subregion CV complexity in this paper.
It is a tractable problem to study subregion CA complexity.
It is expected that the divergent part of the subregion complexity shows a similar behaviour.
However, the universal part of the CV complexity is different from that of the CA complexity in AdS$_3$ \cite{Chapman:2018bqj,Sato:2019kik,Braccia:2019xxi} and it is natural to expect that this observation still holds for the subregion complexities.

There are several proposals for the definition of the complexity in conformal field theory \cite{Jefferson:2017sdb,Chapman:2017rqy,Caputa:2017urj,Caputa:2017yrh,Caputa:2018kdj,Belin:2018bpg}.
It is intriguing to apply these proposals to the moving mirror model.
For this purpose, it is required to generalise these proposals to time-dependent states and BCFT.\footnote{The  generalisation of the path-integral optimization \cite{Caputa:2017urj,Caputa:2017yrh} to BCFT has been discussed in \cite{Sato:2019kik}.}
In particular, the complexity of time-dependent states is an interesting topic. As far as we know, such a situation has not been discussed.

\acknowledgments

The author is grateful to T.~Takayanagi for fruitful discussion and comments on the draft version of this paper.
The author is also grateful to Y.~Kusuki, M.~Honda, 
K.~Ohmori and Y.~Okuyama for useful conversation.
The work is supported by the National Center of Theoretical Sciences (NCTS).

\appendix

\section{Derivation of (\ref{CV_disconnected})}
\label{app1}

In this appendix, we give a detail derivation of \eqref{CV_disconnected}.

\begin{figure}
\centering
\begin{tikzpicture}[scale=1.50]
   \draw[->,>=stealth,thick] (1.4,0)--(2.5,0) node [right] {$X$};
   \draw[thick] (-1,0)--(-0.7,0);
   \draw[red,thick] (-0.7,0)--(1.4,0);
   \draw[->,>=stealth,thick] (0,0)--(0,2) node [above] {$\eta$};
   \draw[thick] (0,0)--(-1,1.732);
   \draw (-1.2,1.732) node [above] {$X = -\alpha \eta$};
   \draw[thick,red] (1.4,0) arc (0:120:1.4);
   \fill (1.4,0) circle (1.2pt) node[below] {$X_\text{R}$} (-0.7,1.732*0.7) circle (1.2pt);
   \draw[thick,dotted,red] (-0.7,1.732*0.7) -- (-0.7,0);
   \filldraw[pattern=north east lines, pattern color=blue,opacity=0.6,draw=none] (0,0)--(-0.7,1.732*0.7)--(-0.7,0)--cycle;
   \draw (0,0) node [below] {$0$};
\end{tikzpicture}
\caption{Entanglement wedge of the semicircle of radius $X_\text{R}$.}
\label{fig:bdy}
\end{figure}
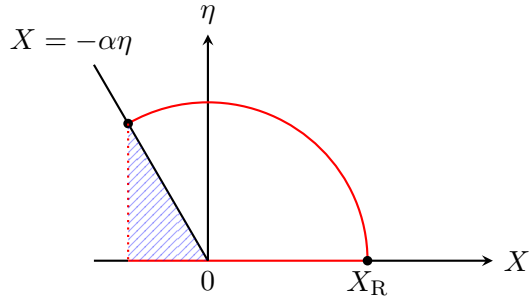

The area of the disconnected phase of the entanglement wedge is equal to the difference between a semicircle of radius $X_\text{R}$ and that of radius $X_\text{L}$.
The area of the semicircle with radius $X_\text{R}$ is given by the difference of the region surrounded by the red line and the triangle painted by the shaded blue. See Fig. \ref{fig:bdy}. 
Considering the tilt of the entanglement wedge, the area of the region surrounded by the red line is given by
\begin{align}
\begin{aligned}
    & \int_{-X_\text{R} \sin \theta}^{X_\text{R}} \! \dd X \int_{\epsilon '}^{\sqrt{1-(T_\text{d}/X_\text{d})^2}\sqrt{X_\text{R}^2-X^2}} \! \dd \eta \frac{L^2}{\eta^2} \sqrt{1-\left(\frac{T_\text{d}}{X_\text{d}}\right)^2} \\
    & = L^2 \int_{-X_\text{R} \sin \theta}^{X_\text{R}} \! \dd X\, \frac{\sqrt{1-(T_\text{d}/X_\text{d})^2}}{\epsilon'} - L^2 \left( \theta +\frac{\pi}{2} \right) \,,
\end{aligned}
\end{align}
and that of the triangle is given by
\begin{align}
\begin{aligned}
        & \int_{-X_\text{R} \sin \theta}^{0} \! \dd X \int_{\epsilon '}^{-\sqrt{1-(T_\text{d}/X_\text{d})^2}X/\tan \theta } \! \dd \eta \frac{L^2}{\eta^2} \sqrt{1-\left(\frac{T_\text{d}}{X_\text{d}}\right)^2} \\
    & = L^2 \int_{-X_\text{R} \sin \theta}^{0} \! \dd X \left( \frac{\sqrt{1-(T_\text{d}/X_\text{d})^2}}{\epsilon'} + \frac{\tan \theta}{X} \right) \,,
\end{aligned}
\end{align}
with $\tan \theta = \alpha$ and $\epsilon' = \epsilon \sqrt{p' (p^{-1}(U))}$.
Since the area is tilting, the Lorentz factor appears both in the upper bound of the radial direction and the determinant of the induced metric.
In total, the area of the arc with the radius $X_\text{R}$ is given by
\begin{align}
\begin{aligned}
    L^2 \int_{0}^{X_\text{R}} \dd X \frac{\sqrt{1-(T_\text{d}/X_\text{d})^2}}{\epsilon'} - L^2 \left( \theta +\frac{\pi}{2} \right) 
    - L^2 \int_{-X_\text{R} \sin \theta}^{0} \! \dd X  \frac{\tan \theta}{X} \,.
\end{aligned}
\end{align}
By subtracting the area with radius $X_\text{L}$, we reproduce the result \eqref{CV_disconnected}.

\bibliographystyle{JHEP}
\bibliography{ComplexityBCFT}

\end{document}